\documentclass[preprint]{aastex}
\usepackage{graphicx}
\usepackage{natbib}

\newcommand{\sles}{\lower2pt\hbox{$\buildrel {\scriptstyle <}\over {\scriptstyle\sim}$}}
\newcommand{\sgreat}{\lower2pt\hbox{$\buildrel {\scriptstyle >}\over {\scriptstyle\sim}$}}
\newcommand{\al}{\alpha_{-1}}
\newcommand{\md}{\dot m}
\newcommand{\mdin}{\dot M_{\rm in}}
\newcommand{\mdout}{\dot M_{\rm out}}
\newcommand{\csq}{c_s^2}
\newcommand{\rout}{r_{\rm out}}
\newcommand{\Too}{T_{11}}
\newcommand{\rhoout}{\rho_{\rm out}}
\newcommand{\qeN}{q_{eN}^-}
\newcommand{\qnu}{q_{\nu\bar\nu}^-}
\newcommand{\tacc}{t_{\rm acc}}

\newcommand{\pgas}{p_{\rm gas}}

\newcommand{\msun}{M_\odot}
\newcommand{\macc}{m_{\rm acc}}
\newcommand{\mdinj}{\dot m_{\rm inj}}

\begin{document}

\title{Accretion Models of Gamma-Ray Bursts}

\author{Ramesh Narayan$^1$, Tsvi Piran$^2$ and Pawan Kumar$^3$}

\affil{1. Harvard-Smithsonian Center for Astrophysics,
Cambridge, MA 02138, USA \\
2. Racah Institute of Physics, Hebrew University,
  Jerusalem 91904, Israel \\
3. Institute for Advanced Studies, Princeton, NJ 08540, USA.}

\begin{abstract}

Many models of gamma-ray bursts (GRBs) involve accretion onto a
compact object, usually a black hole, at a mass accretion rate of
order a fraction of a solar mass per second.  If the accretion disk is
larger than a few tens or hundreds of Schwarzschild radii, the
accretion will proceed via a convection-dominated accretion flow
(CDAF) in which most of the matter escapes to infinity rather than
falling onto the black hole.  Models involving the mergers of black
hole white dwarf binaries and black hole helium star binaries fall in
this category.  These models are unlikely to produce GRBs since very
little mass reaches the black hole.  If the accretion disk is smaller,
then accretion will proceed via neutrino cooling in a
neutrino-dominated accretion disk (NDAF) and most of the mass will
reach the center.  Models involving the mergers of double neutron star
binaries and black hole neutron star binaries fall in this category
and are capable of producing bright GRBs.  If the viscosity parameter
$\alpha$ in the NDAF has a standard value $\sim0.1$, these mergers can
explain short GRBs with durations under a second, but they are
unlikely to produce long GRBs with durations of tens or hundred of
seconds.  If the accretion disk is fed by fallback of material after
a supernova explosion, as in the collapsar model, then the time scale
of the burst is determined by fallback, not accretion.  Such a model
can produce long GRBs.  Fallback models again require that the
accretion should proceed via an NDAF rather than a CDAF in order for a
significant amount of mass to reach the black hole.  This condition
imposes an upper limit on the radius of injection of the gas.

\end{abstract}

\section { Introduction}

The fireball model \citep[see][for reviews]{P99,P00} provides a good
understanding of conditions within the $\gamma$-ray-emitting and
afterglow-emitting regions of gamma-ray bursts (GRBs).  According to
this model, GRBs are produced when relativistic ejecta from a
``central engine'' are slowed down by interactions, either with an
external medium (the external shock model) or among different layers
within the ejecta themselves (the internal shock model).  In the
interactions the kinetic energy in the ejecta is converted to
relativistic electrons which produce the observed radiation.  Among
the many successes of the model we note the observational confirmation
of relativistic motion in the afterglow \citep{Frail97,KP97}.

Despite the successes of the fireball model, the nature of the central
engine remains a mystery.  The problem is that the central engine is
hidden from view; no radiation (apart from gravitational radiation and
neutrinos that may possibly be detected in the distant future) reaches
the observer directly from the engine.  For a number of so-called
``long bursts'' accurate positions have been determined through
observations of their afterglows.  Based on this, there is
circumstantial evidence that these bursts are associated with
star-forming regions (e.g. Bloom et al. 2000).  There is no
information at present on the other class of GRBs, the so-called
``short bursts.''

Although we lack direct evidence on the nature of the central engine,
it is nevertheless widely accepted that GRBs are the result of
cataclysmic events involving either neutron stars or stellar-mass
black holes.  The arguments in support of this hypothesis are
straightforward.  (i) Since bursts radiate the bulk of their energy in
the $\gamma$-ray band, it seems likely that a relativistic object is
behind their production.  (ii) The energy budget ($\sim 10^{51}$ erg)
is comparable to the kinetic energy of ejecta in a supernova
explosion. (iii) Most long bursts are highly variable in gamma-rays,
and so are many short bursts (Nakar \& Piran, 2001a).  In particular,
the ratio of the total duration of the burst to the variability time
scale is large, from which one concludes that the gamma-rays must be
produced in internal shocks \citep{SP97}.  A key feature of internal
shocks is that the observed gamma-ray variability reflects the
variability in the activity of the central engine \citep{SP97}. Since
variability time scales as short as a millisecond are observed, the
engine must contain a compact object of no more than a few solar
masses (otherwise the light-crossing time would exceed the variability
time).

An interesting clue to the nature of GRBs is provided by the durations
of bursts.  While the fastest variability time scale is under a
millisecond, burst durations are usually very much longer.  Long
bursts have durations ranging from 10--1000 seconds, and even short
bursts have a median duration of about 0.3 seconds.  Clearly, whatever
is the physical mechanism behind GRB production, it acts on a much
longer time scale than the fastest dynamical time of the central
engine.

Narayan, Paczy\'nski \& Piran (1992) suggested that the central engine
in GRBs involves the accretion of matter onto a compact star, and that
the energy in the burst is provided by the gravitational energy
released by the accreting gas.  In such a model, the duration of the
burst is set by the viscous time scale of the accreting gas.  In most
accretion flows, the viscous time is significantly longer than the
dynamical time, and so the accretion model naturally explains the
large difference between the durations of bursts and their fastest
variability time.

The formation of an accretion disk is a natural outcome of most
popular models of GRBs, e.g. the mergers of double neutron star
binaries (Eichler et al., 1989; Narayan, Paczy\'nski \& Piran
1992), neutron star black hole binaries (Paczy\'nski, 1991;
Narayan Paczy\'nski \& Piran 1992), black hole white dwarf
binaries (Fryer et al., 1999), black hole helium star binaries
(Fryer \& Woosley, 1998), and models based on ``failed
supernovae'' or ``collapsars'' (Woosley, 1993; Paczynski, 1998;
MacFadyen \& Woosley, 1999). An important exception is Usov's
(1992) model in which the GRB energy is provided by the magnetic
and rotational energy of a newly formed rapidly rotating neutron
star.

In this paper, we consider a generic accretion model of a GRB in which
a certain amount of mass, $m_d\msun$, goes into orbit around a
relativistic star of mass $3m_3\msun$.  We assume that the orbiting
mass is initially inserted into a torus at a radius $\rout R_S$, where
$R_S$ is the Schwarzschild radius of the central star: $ R_S={2GM/
c^2}=8.85\times10^5m_3 ~{\rm cm}$.  Starting from the initial toroidal
configuration, the mass spreads out by viscosity and becomes an
accretion flow extending from $r=1$ (the horizon of the central black
hole) to $r\sim\rout$.  (There might also be an outflow at radii
$>\rout$ as we discuss below.)  We work out in this paper the time
scale, $\tacc$, of the accretion flow, the average mass accretion
rate, $\dot M=\md\msun{\rm s^{-1}}$, onto the central star, the amount
of mass, $\macc=\tacc\md$, accreted by the star, and the accretion
efficiency, $\xi=\macc/m_d$.

These parameters are constrained by observations.  In binary merger
models, the durations of bursts should be comparable to the accretion
time $\tacc$.  Durations are of order a second or less for the class
of ``short duration GRBs'' and in the range 10--1000 seconds for the
class of ``long duration GRBs.''  The energy in long duration bursts
is estimated to be $\sim5\times10^{50}$ erg (Panaitescu \& Kumar,
2001; Frail et al. 2001).  With a reasonable efficiency of converting
accretion energy to relativistic flow ($\sles\ 0.01$) this corresponds
to $\macc\ \sgreat\ 0.1 M_\odot$ and to a peak accretion rate $\dot M\
\sgreat\ 10^{-2} M_\odot{\rm s^{-1}}$.  The values of these parameters
for short bursts are less certain, as the distance scale to this
population is uncertain. However, it is unlikely that they are smaller
by more than an order of magnitude.

The mass accretion rate tends to be extremely high for a typical GRB
model --- it is of order a fraction of a solar mass per second.  At
such accretion rates, the optical depth of the accreting gas is
enormous and radiation is trapped inside the gas.  In the normal
course, the accretion would proceed via a radiatively inefficient
flow, such as an advection-dominated accretion flow or the related
convection-dominated accretion flow.  We consider such flows in \S{2}.
If the accreting gas has a sufficiently high temperature and density,
however, it can cool via neutrino emission, leading to a
neutrino-dominated accretion flow, as discussed by Popham, Woosley \&
Fryer (1999) (see also Ruffert \& Janka, 1999, for a numerical
simulation of a binary merger which included neutrino losses).  We
discuss this case in \S{3}. In \S{4} we go beyond the analytical
results of \S\S{2,3} and present numerical results, delineating the
regions of $(\rout,\,m_d)$ space where different types of accretion
occur. We discuss in \S{5} the implications of the results for various
models of GRBs.

\section{Radiatively Inefficient Accretion: CDAF}

At the mass accretion rates of interest to us, the optical depth of
the gas is extremely high, and the radiation is very effectively
trapped (we demonstrate this below).  The accretion flow then
corresponds to an advection-dominated accretion flow, or ADAF (Narayan
\& Yi 1994, 1995; Abramowicz et al. 1995; Chen et al. 1995; see
Narayan, Mahadevan \& Quataert 1998 and Kato, Fukue \& Mineshige 1998
for reviews).  Radiation-trapped ADAFs were initially discussed by
Katz (1977) and Begelman (1978), and later analyzed via
height-integrated ``slim disk equations'' by Abramowicz et al. (1988).

Igumenshchev and co-workers have carried out hydrodynamic simulations
of ADAFs and have discovered several interesting properties of these
flows (Igumenshchev, Chen \& Abramowicz 1996; Igumenshchev \&
Abramowicz 1999, 2000).  They find that, when the dimensionless
viscosity parameter $\alpha$ is large, say $\alpha\ \sgreat\ 0.3$, the
accretion flow has a strong bipolar outflow, as anticipated in some
previous papers (Narayan \& Yi 1994, 1995; Blandford \& Begelman
1999).  On the other hand, when the viscosity is relatively weak, say
$\alpha\ \sles\ 0.1$, the flow has well-developed convection (which
was again anticipated in earlier work, cf. Narayan \& Yi 1994, 1995).
A convective ADAF has quite unusual properties (Stone, Pringle \&
Begelman 1999; Narayan, Igumenshchev \& Abramowicz 2000; Quataert \&
Gruzinov 2000; Igumenshchev, Abramowicz \& Narayan 2000), which arise
because convection moves angular momentum inward rather than outward
(Narayan et al. 2000; Quataert \& Gruzinov 2000).  Convective ADAFs
have been given the name of convection-dominated accretion flows, or
CDAFs.

As in other accretion flows, ``viscosity'' in radiation-trapped flows
is believed to arise from magnetic stresses resulting from the
Balbus-Hawley instability \citep{BH91}.  Numerical simulations of
shearing MHD flows give values of $\alpha$ in the range $\alpha\
\sles\ 0.1$.  We therefore assume that radiation-trapped accretion
flows in GRBs also have $\alpha$ in this range (no 3D MHD simulations
of these flows have been done so far, but there has been some 2D work
by Stone \& Pringle 2001).  Given the relatively low value of
$\alpha$, we expect the accretion flow to take the form of a CDAF.

Ball, Narayan \& Quataert (2001) have given approximate scalings for
various fluid variables in a CDAF.  The density and velocity scale as
\begin{eqnarray}
\rho(r) &\approx &\rhoout\left({r\over\rout}\right)^{-1/2}=
2.97\times10^{14}m_3^{-3}m_d\rout^{-5/2}r^{-1/2} ~{\rm g\,cm^{-3}},
\nonumber
\\
v(r)&\approx& cr^{-3/2}=3\times10^{10}r^{-3/2},
\end{eqnarray}
where, in the first equation, we have used $4\pi \rout^3R_S^3({H/ R})
\rhoout=m_d\msun$ to relate $\rhoout$ in the CDAF to the mass in the
accretion flow:
\begin{equation}
\rhoout = 2.97\times10^{14}m_3^{-3}m_d\rout^{-3} ~{\rm g\,cm^{-3}}.
\end{equation}
Note the use of $4\pi$ rather than $4\pi/3$ in the equation for
the mass.  This has been done to obtain a better match with the
results for an NDAF (see \S{3}).

We assume that the accreting gas consists of photo-disintegrated
nuclei with roughly equal numbers of neutrons and protons (and
electrons).  The optical depth through the flow is
\begin{equation}
\tau_{\rm out}\approx {\rhoout\over2m_p}\sigma_T\rout R_S
=5.21\times10^{19}m_3^{-2}m_d\rout^{-2}.
\end{equation}
The diffusion time for radiation to leak out of the flow is then
\begin{equation}
t_{\rm diff}\approx\tau_{\rm out}{\rout R_S\over c}
=1.54\times10^{15}m_3^{-1}m_d\rout^{-1} ~{\rm s}.
\end{equation}
This time is very much longer than the accretion time for the
parameter ranges of interest to us, namely $m_3\sim1$, $m_d\sim0.1-1$,
$\rout\sim10-10^4$.  Thus we expect radiation to be very effectively
trapped within the accretion flow.

Ball et al. (2001) also estimate the isothermal sound speed $c_s$ and
the scale height $H$ of the accretion flow.  We use slightly different
coefficients here which are more appropriate for a radiation-pressure
dominated $\gamma=4/3$ gas (as opposed to the $\gamma=5/3$ gas that
Ball et al. 2001, considered):
\begin{eqnarray}
\csq&\approx& 0.3c^2r^{-1}=2.70\times10^{20}r^{-1} ~{\rm cm^2\,s^{-2}},
\nonumber \\
{H\over R}&=&{c_s\over\Omega_K R}\approx 0.77.
\end{eqnarray}
These scalings are approximate, but they are likely to be accurate
enough for the purposes of this paper.  The mass accretion rate in the
CDAF is estimated to be $\dot M=4\pi r^2R_S^2({H/ R})\rho(r)v(r)$,
which gives, using the above relation for $H/R$,
\begin{equation}
\md=3.39\times10^4m_3^{-1}m_d\rout^{-5/2}.
\end{equation}
(Recall that $\md$ is defined as the mass accretion rate in units of
solar masses per second.)

The accretion time scale is not simply equal to $m_d/\md$.  The reason
is that much of the mass in a CDAF actually flows out of the system
rather than into the central black hole (Stone et al. 1999; Narayan et
al. 2000; Quataert \& Gruzinov 2000Igumenshchev \& Abramowicz 2000).
We therefore proceed as follows.

The random velocities of convective blobs in a CDAF are typically less
than the local Keplerian velocity by a factor which is proportional to
the viscosity parameter $\alpha$ (see Narayan et al. 2000). We thus
write approximately
\begin{equation}
v_{\rm turb}\approx\alpha v_K=2.12\times10^9\al r^{-1/2}
~{\rm cm\,s^{-1}}.
\end{equation}
The residence time of a convective blob at radius $r$ is then
\begin{equation}
t_{\rm res}\approx{rR_S\over v_{\rm turb}}=4.17\times10^{-4}
\al^{-1}m_3r^{3/2} ~{\rm s}.
\end{equation}
We make the reasonable assumption that the accretion time is of
the same order as $t_{\rm res}$ at $r=\rout$:
\begin{equation}
\tacc\approx t_{\rm res}(\rout)=4.17\times10^{-4}
\al^{-1}m_3\rout^{3/2} ~{\rm s}.
\end{equation}
We then find that the amount of mass accreted by the black hole is
$\macc=\tacc\md$, but since this mass cannot exceed the total
available mass $m_d$ we write:
\begin{eqnarray}\macc&=&m_d, \qquad \qquad \qquad  \rout\leq 14.1\al^{-1}m_d,
\nonumber
\\
&=&14.1\al^{-1}m_d\rout^{-1}, ~~~\rout>14.1\al^{-1}m_d.
\end{eqnarray}

Note that, when $\rout$ is large, the accreted mass is much less than
$m_d$.  The reason is that the bulk of the mass is ejected from the
system, flowing out at $r\sim\rout$.  The energy for the ejection is
provided by convective energy flux from the interior of the flow.

The coefficient 14 in equation (10) is somewhat uncertain since we do
not know the exact relation between $\tacc$ and $t_{\rm res}$; there
could well be a numerical factor other than unity relating the two.  A
different (and more more detailed) derivation of equation (10) is
given in the Appendix, where we again find that the coefficient is
uncertain.  In the rest of the paper we use equation (10) as written,
but we should keep in mind that it could be in error by a factor of a
few.

Let us now calculate the temperature of the CDAF.  The pressure in the
accreting gas is given by
\begin{equation}
p=\rho\csq=8.02\times10^{34}m_3^{-3}m_d\rout^{-5/2}r^{-3/2}
~{\rm erg\,cm^{-3}}.
\end{equation}
The pressure has three contributions: radiation pressure, gas
pressure, and degeneracy pressure (Popham et al. 1999):
\begin{equation}
p={11\over12}aT^4+{\rho kT\over m_p}+ { 2 \pi h c \over 3} \left({3 \over 8 \pi m_p}
\right)^{4/3}\left({\rho \over \mu_e}\right)^{4/3}.
\label{pressure}
\end{equation}
The quantity $a$ is the radiation constant, and the factor 11/12
includes the contribution of relativistic electron-positron pairs
(asssuming that the temperature is sufficiently above the pair
threshold limit).  The gas pressure term includes the
contributions from non-relativistic particles.  We assume that we
have an equal mix of protons and neutrons (i.e. we assume that
all complex nuclei have been photo-disintegrated).  In the
degeneracy pressure term we use an electron molecular weight
$\mu_e=2$, assuming an equal mix of protons and neutrons; the
ratio of neutrons to protons in the interior of a NS is about
0.2.  The contribution of $e^-$--$e^+$ pairs has been ignored in
the degeneracy pressure term.  For the calculations presented in
\S{4}, we solve for $T$ by using the full expression for $p$
given in equation (\ref{pressure}).  But here, in order to obtain
simple analytic estimates, we simplify the equation and assume
that radiation pressure dominates.  We then find that
\begin{equation}
T=1.84\times10^{12}m_3^{-3/4} m_d^{1/4}\rout^{-5/8}r^{-3/8}~K.
\end{equation}
This estimate of $T$ is valid only if gas pressure is smaller than
radiation pressure.  The gas pressure is equal to
\begin{equation}
\pgas={\rho kT\over m_p}=4.52\times10^{34}m_3^{-15/4}m_d^{5/4}
\rout^{-25/8}r^{-7/8}~{\rm erg\,cm^{-3}}.
\end{equation}
For this to be smaller than radiation pressure we require
\begin{equation}
{r\over\rout}<1.5m_3^{6/5}m_d^{-2/5}.
\end{equation}
For our fiducial black hole of mass $3\msun$, gas pressure
dominates only if the disk is quite massive, $m_d\ \sgreat\ 5$,
and if $r$ is close to $\rout$.  (The detailed numerical
calculations in \S{4} show that gas pressure dominates for $m_d\
\sgreat\ 1$ rather than 5, but even for such values of $m_d$, if
$r$ is much less than $\rout$, radiation pressure takes over.)

Let us next calculate the rate of cooling of the accreting gas.  We
showed earlier that cooling via radiative diffusion is negligible.
However, at the temperatures found in these flows, cooling via
neutrino losses may be important (cf. Popham et al. 1999).  The
cooling rate per unit volume due to neutrinos, $q^-$, takes the form
\begin{equation}
q^-=\qnu+\qeN \approx 5\times10^{33}\Too^9 +
9.0\times10^{23}\rho \Too^6 ~{\rm erg\,cm^{-3}\,s^{-1}} ,
\label{q-}
\end{equation}
where the first term on the right describes cooling via pair
annihilation (the so-called URCA process) and the second term
describes cooling via pair capture on nuclei (estimated for $X_{\rm
nuc}=1$ as appropriate for our fully photo-disintegrated nuclear gas,
cf. Popham et al. 1999).  For the range of parameters of interest to
us, $\qeN$ invariably dominates over $\qnu$.  We therefore neglect
$\qnu$ hereafter.

We then obtain the following estimate for the cooling rate in the
CDAF:
\begin{equation}
\qeN=1.04\times10^{46}m_3^{-15/2}m_d^{5/2}\rout^{-25/4}r^{-11/4}
~{\rm erg\,cm^{-3}\,s^{-1}}.
\end{equation}
Since the energy density of the radiation-dominated gas is equal to
$3p_{\rm rad}=(11/4)aT^4$, we estimate the cooling time of the gas at
radius $r$ to be
\begin{equation}
t_{\rm cool}={3p_{\rm rad}\over\qeN}=
2.31\times10^{-11}m_3^{9/2}m_d^{-3/2}\rout^{15/4}
r^{5/4} ~{\rm s}.
\end{equation}
One of the important requirements for the accretion flow to behave
like a CDAF is that it should be radiatively inefficient.  We,
therefore, require the cooling time $t_{\rm cool}$ at each radius to
be longer than the residence time $t_{\rm res}$ of a convecting blob
at that radius.  The ratio of cooling to residence time is
\begin{equation}
{t_{\rm cool}\over t_{\rm res}}=5.54\times10^{-8}\al
m_3^{7/2}m_d^{-3/2}\rout^{15/4}r^{-1/4}.
\end{equation}
Since the ratio decreases with increasing $r$, we set $r=\rout$.  Then
we see that a CDAF is possible only if the following condition is
satisfied:
\begin{equation}
\rout>118\al^{-2/7}m_3^{-1}m_d^{3/7}.
\label{CDAF}
\end{equation}
If the mass in the accretion flow is initially at a radius greater
than the above limit, then the flow will become a CDAF at $r=\rout$,
and indeed for all $r<\rout$.  The scaling relations derived in this
section would then be valid.  If, however, the initial radius of the
gas is below the above limit, then neutrino-cooling will prevail, and
we will have a cooling-dominated accretion flow at all radii from
$\rout$ down to the black hole.

For completeness, we consider also the case when gas pressure
dominates (which requires a large value of $m_d$, and $r$ close to
$\rout$, as shown above).  The temperature is then obtained by setting
$p=\pgas$.  This gives
\begin{equation}
T={m_p\csq\over k}=3.27\times10^{12}r^{-1} ~K.
\end{equation}

The analysis presented so far assumes that we are given the initial
mass in the disk $m_d$ and the initial radius of the gas $\rout$.
Another possibility is that we have steady injection of mass at a rate
$\mdinj\msun{\rm s^{-1}}$ at a circularization radius $\rout$.  This
is the case, for instance, in the collapsar model, where the material
is supplied by fallback from a supernova explosion.  We may assume
that the accretion flow achieves an approximate steady state in which
the rate of injection of mass equals the rate of mass loss from the
flow (both inflow and outflow): $\mdinj=m_d/\tacc$.  Therefore, we
obtain
\begin{equation}
m_d=4.17\times10^{-4}\al^{-1} m_3\mdinj\rout^{3/2}.
\label{minj}
\end{equation}
This expression, which gives the mapping between $m_d$ and $\mdinj$,
may be substituted into the various relations derived in this section
to obtain the corresponding results for the case of steady mass
injection.

\section{Radiatively Efficient Accretion: NDAF}

We saw in the previous section that when the mass is initially
injected at an outer radius smaller than the limit given in equation
(20), cooling via neutrino emission becomes significant and the flow
is no longer radiatively inefficient.  We then have a regime of
accretion in which the viscous energy dissipation is balanced by
neutrino cooling.  Popham et al. (1999) named this a
neutrino-dominated accretion flow (NDAF) and worked out its
properties.  We extend their results in this section.

Because the gas cools efficiently, an NDAF behaves like a thin
accretion disk and we are entitled to use the basic theory of thin
disks (Shakura \& Sunyaev 1973; Frank, King \& Raine 1992).  As
before, the pressure of the gas has three contributions and is
described by equation (\ref{pressure}).  The isothermal sound speed
$c_s$ and the vertical scale height $H$ are given by $\csq={p/\rho}$
and $ H={c_s/\Omega_K}$, where $\Omega_K = ({GM/
R^3})^{1/2}=2.40\times10^4m_3^{-1}r^{-3/2} ~{\rm \ s^{-1}}$ is the
Keplerian angular velocity.  We write the coefficient of kinematic
viscosity in the usual form as
\begin{equation}
\nu=\alpha{\csq\over\Omega_K},
\end{equation}
where $\alpha$ is a dimensionless parameter which we expect to have a
value $\sim0.1$.  We denote, therefore, $\al = \alpha/0.1$, and write
down our scalings in terms of $\al$.

A thin accretion disk satisfies two equations.  Angular momentum
balance gives
\begin{equation}
\dot M=3\pi\nu\Sigma\left[1-\left({R\over R_*}\right)^{-1/2}\right]
\approx 6\pi\nu\rho H,
\end{equation}
where $\Sigma=2\rho H$ is the surface density and $R_*$ is the radius
of the inner edge of the disk ($3R_S$ for a Schwarzschild black hole).
The approximation on the right is valid for $R\gg R_*$.  The radial
velocity of the gas is given by $v=3\nu/2R$ and so the accretion time
is
\begin{equation}
\tacc={R_{\rm out}\over v(R_{\rm out})}=
{2R_{\rm out}^2\over 3\nu}.
\label{tacc}
\end{equation}
If we write the disk mass as $m_d=\tacc\dot M$, then by combining the
previous two equations we find that
\begin{equation}
m_d=4\pi R_{\rm out}^3\left({H\over R}\right)_{\rm out}\rhoout.
\end{equation}
Note that we used this relation, with the same coefficient $4\pi$,
also for a CDAF.

The condition of energy balance (viscous heating equals radiative
losses) gives:
\begin{equation}
{3GM\dot M\over 8\pi R^3}=q^- H=(\qeN+\qnu)H.
\end{equation}
This relation closes our set of equations and allows us to solve
for $\rho$, $T$ and other quantities.  We show numerical
solutions of the full equations in \S{4}.

In the rest of this section we derive analytical scalings by making
some approximations.  First, the numerical calculations show that the
$\qeN$ term in the cooling law dominates over $\qnu$ for all
parameters of interest.  We will therefore assume this.  The
calculations also show that for most parameters, either gas pressure
or degeneracy pressure dominates.  (There is a small region of
parameter space where radiation pressure dominates; we ignore this
region in the analytical work.)

Let us first assume that gas pressure dominates.  Expressing all
results in terms of the scaled temperature, $\Too=T/10^{11}K$, and
substituting in the angular momentum equation, we obtain a relation
between $\rho$ and $\Too$
\begin{equation}
\rho=2.56\times10^{13}\al^{-1}m_3^{-2}\md r^{-3}\Too^{-3/2} ~{\rm g\,cm^{-3}}.
\end{equation}
Substituting this in the energy equation, we can solve for $\Too$, and
thereby obtain the various other quantities (Popham et al. have
derived similar relations):
\begin{eqnarray}
\Too&=&0.548\al^{1/5}m_3^{-1/5}r^{-3/10},
\nonumber \\
\tacc&=&2.76\times10^{-2}\al^{-6/5}m_3^{6/5}\rout^{4/5}
~{\rm s},
\nonumber \\
\md&=&36.2\al^{6/5}m_3^{-6/5}m_d\rout^{-4/5},
\\
\rhoout&=&2.28\times10^{15}\al^{-1/10}m_3^{-29/10}m_d\rout^{-67/20}
~{\rm g\,cm^{-3}}.
\nonumber
\end{eqnarray}
In a cooling-dominated thin disk, very little mass is expected to be
lost to outflows, so we expect nearly all the mass in the disk to be
accreted by the star, i.e.
\begin{equation}
\macc\approx m_d.
\end{equation}

The above results are valid provided gas pressure dominates.  By
comparing gas pressure to degeneracy pressure we can determine the
condition for this to be true.  We find that we require
\begin{equation}
\rout>26.2\al^{-2/7}m_3^{-46/49}m_d^{20/49}.
\label{gas}
\end{equation}
When the outer radius is below this limit, degeneracy pressure takes
over from gas pressure and we obtain a different set of analytical
scalings:
\begin{eqnarray}
\md&=&355\al m_3^{-13/7}m_d^{9/7}\rout^{-3/2},
\nonumber \\
\rhoout&=&7.32\times10^{14}m_3^{-18/7}\md^{6/7}\rout^{-3} ~{\rm g\,cm^{-3}}.
\nonumber \\
\Too&=&0.800\al^{1/6}m_3^{-13/42}m_d^{1/21}\rout^{-5/12},
\\
\tacc&=&2.82\times10^{-3}\al^{-1}m_3^{13/7}m_d^{-2/7}\rout^{3/2} ~{\rm s}.
\nonumber
\end{eqnarray}

In our analysis of both the gas-pressure-dominated and
degeneracy-pressure-dominated regimes, we assumed that the accreting
gas is optically thin to its own neutrino emission.  This assumption
breaks down at sufficiently small radii.  We may estimate the neutrino
optical depth as
\begin{equation}
\tau_\nu={\qeN H\over 4\sigma T^4}=5.20\times10^3\al^{-2/3}
m_3^{-55/21}m_d^{23/21}\rout^{-17/6}.
\end{equation}
The radius at which the optical depth goes to unity is
\begin{equation}
r_{\tau=1}=20.5\al^{-4/17}m_3^{-110/119}m_d^{46/119}.
\label{opt_depth}
\end{equation}
This radius lies within the degeneracy-pressure-dominated zone.
Inside this radius, we need to consider neutrino transport in more
detail.  This is beyond the scope of the paper.

It is interesting to examine whether the NDAF solution is stable.
Following Piran (1978) we use the general condition for thermal
stability:
\begin{equation}
\left({d \ln Q^+ \over d \ln H}\right)_{|\Sigma} < \left({d \ln Q^-
\over d \ln H}\right)_{|\Sigma} \ ,
\label{stability}
\end{equation}
where $Q^\pm$ are the integrated (over the height of the disk) heating
($+$) and cooling ($-$) rates.  For an NDAF with viscosity described
by the $\alpha$-prescription, the heating rate goes as $Q^+ \propto
H^2 $, and the vertically integrated cooling rate (for pair capture on
nuclei) goes as $Q^- \propto \Sigma T^6$.
For an NDAF in which gas pressure dominates, $T \propto H^2$ and $Q^-
\propto \Sigma H^{12}$.  The criterion (\ref{stability}) is satisfied
easily and we conclude that such an NDAF is thermally stable.

If radiation pressure dominates, $T^4 \propto p \propto \Sigma H$ and
$Q^- \propto \Sigma^{5/2} H^{3/2}$.  The condition (\ref{stability})
is not satisfied and the NDAF is unstable. This resembles the
situation in ``conventional'' accretion disks whose inner regions become
thermally unstable when radiation pressure dominates (Frank et
al. 1992).

If degeneracy pressure dominates, the temperature is independent of
either $H$ or $\Sigma $.  The temperature of the disk can then freely
adjust such that $Q^-$ balances any $Q^+$.  The situation is clearly
thermally stable.

The condition for viscous stability is:
\begin {equation}
{d \dot M  \over d \Sigma} > 0 .
\end{equation}
We have $\dot M \propto \Sigma$ for the gas pressure case, $\dot M
\propto \Sigma^7$ for the radiation pressure case, and $\dot M \propto
\Sigma^{9/7}$ for the degeneracy pressure case.  All three cases are
viscously stable.

Thus, combining all these results on stability, we find that an
optically thin NADF is unstable only if it is radiation pressure
dominated.  There is a very narrow region of parameter space near
the boundary between NDAFs and CDAFs in $(\rout,\md)$ space where
radiation pressure does dominate. This unstable region could
conceivably play a role in determining the temporal behavior of
some bursts. The stability properties of the optically thick
regions of the NDAF remain to be worked out.


The disks that we are considering are massive and we should also
consider gravitational instabilities.  The CDAF zone is always
gravitationally stable in our models; the Toomre $Q$ parameter is
invariably much greater than unity.  For gas pressure dominated NDAFs,
we find that the Toomre $Q$ parameter is given by
\begin{equation}
Q = 8.5 ~\al^{1/10} m_3^{9/10} m_d^{-1} r^{-9/20} r_{out}^{4/5}.
\end{equation}
We see that $Q$ decreases with increasing $r$, so that the flow is
most unstable on the outside.  However, even for $r = r_{out}$, $Q$ is
sufficiently larger than unity (for all reasonable values of
parameters) that we are guaranteed stability.  A similar result
applies to radiation pressure dominated NDAFs.

The case of degeneracy pressure dominated NDAFs is more
complicated. For most of the parameter space $(m_{d}, r_{out})$ we
find $Q$ to be greater than unity, but it is only marginally so
($Q\approx 3$).  When $r_{out}<10$, we find that $Q$ might become less
than unity, signifying gravitational instability.  But this is also
the region in which the disk becomes optically thick to neutrinos and
the analysis we have carried out breaks down.


\section{Numerical Results}

We discuss in this section numerical results which we have
obtained by solving the full equations.  We assumed that we are
given the initial mass of the accretion disk $m_d$ and the
initial radius $\rout$.  For each choice of $\rout$ and $m_d$, we
first assumed that the flow consists of a CDAF and used the
relations described in \S{2} to calculate the properties of the
accreting gas.  In particular, we used equations (11) and
(\ref{pressure}), with $\rho$ given by equation (1), to solve for
the temperature $T$ as a function of $r$.  We checked the flow at
all radii from $r=\rout$ down to $r=1$ to make sure that the gas
is radiatively inefficient at all radii. Specifically, we
estimated the cooling time $t_{\rm cool}$, using the cooling
formula given in equation (\ref{q-}), and checked that $t_{\rm
cool}$ is longer than the residence time of convective blobs
$t_{\rm res}$ at that radius.  If $t_{\rm cool}>t_{\rm res}$ at
all $r$, then we identified the flow as a pure CDAF and estimated
the accretion time $\tacc$ and the total mass accreted $\macc$,
using the formulae given in \S{2}.

For a small region of parameter space near the transition between a
pure CDAF and a pure NDAF, we found that the flow starts off as a CDAF
at $r=\rout$ and switches to an NDAF at a smaller radius.  We
estimated $\tacc$ and $\macc$ for these cases by using a reasonable
matching formula at the transition radius.  (We do not provide the
details here, since this case is seen for only a small range of
parameters).

For $\rout$ less than a critical value (whose value depends on
$m_d$), we found that the flow cools too rapidly at $r=\rout$ to
be a CDAF. In such cases, the flow becomes an NDAF at $r=\rout$
and remains an NDAF all the way down to $r=1$.  We solved the
corresponding set of equations (see \S{3}), and estimated $\tacc$
and $\macc$ appropriately.

Figures 1 \& 2 show contours of $\tacc$ and the accretion
``efficiency'' $\xi\equiv\macc/m_d$, plotted in the space of the
two principal parameters of the problem, $\rout$ and $m_d$. These
models have the ``canonical values" $m_3=\al=1$.  The boundary
between the CDAF and NDAF zones is clearly seen in both figures,
but especially in Fig. 1.  The numerically determined location of
the CDAF-NDAF boundary is fairly close to the analytical
approximation given in equation (20) and shown as the right-most
dotted line in Fig. 1.

As equation (9) shows, the accretion time scale $\tacc$ for a CDAF is
independent of $m_d$ and depends only on $\rout$.  In contrast,
$\tacc$ has a more complicated dependence on $\rout$ and $m_d$ for an
NDAF (eqs. 29 and 32).  These behaviors are seen clearly in Fig. 1.

For a given initial mass $m_d$, the accreted mass $\macc$ is equal to
$m_d$, independent of $\rout$, in the case of an NDAF, and so $\xi=1$
(see Fig. 2).  However, if the flow becomes a CDAF (which happens for
$\rout$ greater than a critical value), $\macc$ is significantly less
than $m_d$, and the rest of the mass is ejected from the system;
$\macc$ scales as $\rout^{-1}$ in this case (cf eq 10).  Therefore,
$\xi$ becomes significantly less than unity.  Figure 2 illustrates
this dependence.

In some GRB models, such as those involving a BH-NS merger
or the collapse of a very massive star, the mass
of the black hole could be larger than $3M_\odot$.  We have therefore
computed results for an $M=30M_\odot$ black hole, i.e.  $m_3=10$.
Equations (20), (31) and (34) show that the various critical radii
vary roughly inversely as $m_3$ (which is equivalent to saying that
the physical radii $R=rR_S$ at which the corresponding transitions
occur are independent of the black hole mass).  We have confirmed this
result in the detailed numerical calculations.  As a result, for
$m_3=10$, the NDAF zone shrinks by a factor of 10 in $\rout$, and the
various variants of the NDAF (degeneracy pressure dominated zone,
optically thick zone) practically disappear.  Thus, large black hole
masses are not conducive to the formation of an interesting NDAF zone.

In addition, we investigated models with a smaller value of the
viscosity coefficient: $\alpha=0.01$, i.e. $\al=0.1$.  In this case,
the various critical radii become larger by roughly a factor of 2, as
expected from equations (20), (31) and (34).  More importantly,
$\tacc$ increases by a factor of $\sim10$ (see equation \ref{tacc}).

We also considered the case when the accretion is fed at a
constant rate $\mdinj$ (rather than being initiated with an
instantaneous addition of mass $m_d$ as we have assumed so far).
This case could be relevant for collapsar models.  Figure 3 shows
contours of accretion efficiency defined in this case as $\xi =
\dot m/\mdinj$, plotted in the space of $\rout$ and $\mdinj$ for
the ``canonical values" $m_3=\al=1$.  The overall behavior is
consistent with the scaling relations derived in \S\S{2,3}.

\section{Discussion}

The starting point for this work is the fact that the accretion flow
in a putative GRB central engine can have two very different forms: we
expect the accretion to occur as a radiatively inefficient CDAF
(convection-dominated accretion flow, cf Narayan et al. 2000, Quataert
\& Gruzinov 2000) if the mass $m_d$ is introduced at a somewhat large
outer radius, namely $\rout$ greater than the limit given in equation
(20), while we expect the accretion to proceed via a radiatively
efficient NDAF (neutrino-dominated accretion flow, cf. Popham et
al. 1999) if $\rout$ is smaller than this limit.  For a narrow zone in
parameter space close to the CDAF/NDAF transition, it is possible for
the flow to be a CDAF on the outside ($r\ \sles\ \rout$) and to switch
to an NDAF on the inside.  But this is rare.  By and large, for most
choices of $\rout$ and $m_d$, the flow is either a CDAF at all radii
or an NDAF at all radii.  Since the two kinds of flow are very
different from each other, there is a rather large difference in what
an observer would see in the two cases.

In most of the region of $(\rout, m_d)$ space where the flow is a
CDAF, the mass accretion rate $\md$ and the amount of mass accreted
$\macc$ are both very small.  This is because in a CDAF, especially
when $\rout$ is large, much more mass flows out of the system than
into the black hole.  If, as seems reasonable, a GRB engine requires a
relatively large $\macc$ in order to produce a viable burst, then our
work suggests that systems which form CDAFs are less likely to produce
observable bursts.

In contrast, an NDAF has substantially larger values of $\md$ and
$\macc$ for a given $m_d$.  (In fact, we assume that $\macc=m_d$ for
an NDAF; there is probably some mass loss in a wind, but we expect it
to be a small fraction of the total mass.)  Therefore, a source with
an NDAF is a much more plausible model of a GRB engine.  By this
argument, neutrino cooling (the key ingredient of an NDAF) is
important for an efficient GRB.  Note that we are not assuming
anything about the actual mechanism of a burst.  The fireball may be
produced through neutrino-antineutrino annihilation \citep{Eichler89},
or it could be the result of some other mechanism.  Our suggestion is
that whatever the mechanism may be, if it is based on accretion, then
it probably requires a large $\macc$ to operate efficiently; such an
$\macc$ can be achieved only through neutrino cooling of the accreting
gas in an NDAF model.

We note a possible caveat to the above conclusion.  The energy in long
duration GRBs is estimated to be about $5\times 10^{50}$ erg
(Panaitescu \& Kumar, 2001; Frail et al. 2001).  This energy could in
principle be generated in a CDAF model with $\macc\ \sles\
0.1M_\odot$, provided the gravitational energy of the accreted matter
is used very efficiently to launch a high Lorentz factor wind.  We
feel, however, that it is reasonable to hypothesize that the
efficiency for converting the gravitational energy of $\macc$ to a
GRB-producing relativistic wind is low, say not larger than 1\%, and
that bursts arise only from systems that form NDAFs and have large
$\macc$.

\begin{figure}
  \begin{center}
    \includegraphics[width=12cm]{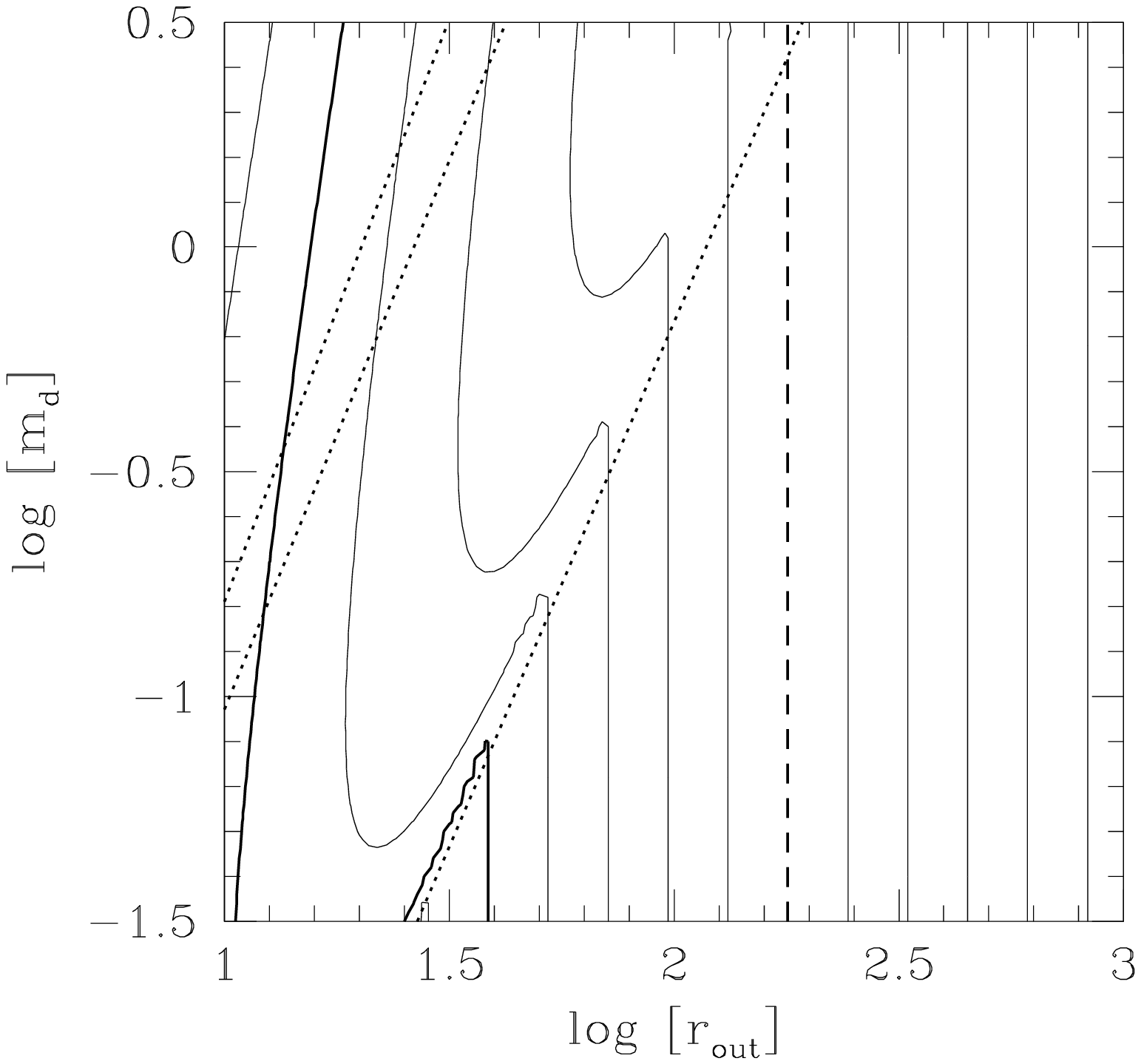}
    \caption{Contours of $\tacc$ in the $\rout$ (in units of $R_S$)
    $m_d$ (in solar masses) plane, for $m_3=1$ and $\al=1$.  The
    contours are equally spaced in log. The lowest contour is at
    $\log({\rm quantity})=-1.2$, and succeeding contours are shifted by
    +0.2 up to a maximum value of +1.  Two contours are highlighted:
    the contour at $-1$ is shown as a bold solid line and the contour
    at 0 is shown as a bold dashed line.  The three dotted lines
    correspond to three important transitions, computed with the
    analytical approximations of \S\S{2,3}.  Starting from the
    right, the lines correspond to: (i) the transition from a CDAF to
    an NDAF (calculated via equation \ref{CDAF}), (ii) the transition
    from a gas-pressure-dominated NDAF to a
    degeneracy-pressure-dominated NDAF (equation \ref{gas}), and (iii)
    the transition from an NDAF that is optically thin to neutrino
    emission to one that is optically thick (equation
    \ref{opt_depth}). Our calculations are not reliable to the left of
    the leftmost dotted line since we have not allowed for optical
    depth effects in the model. } \label{Fig1} \end{center}
\end{figure}

\begin{figure}
  \begin{center}
    \includegraphics[width=12cm]{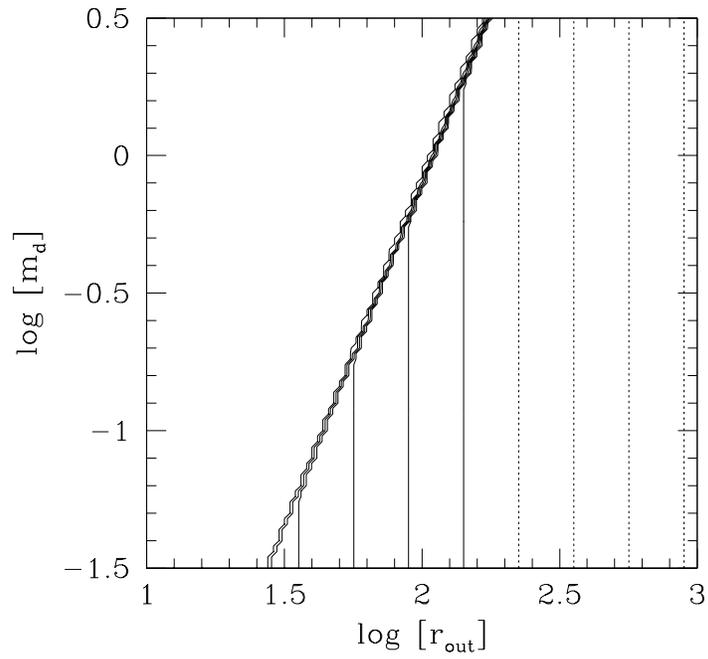}
    \caption{Contours of accretion efficiency $\xi=\macc/m_d$ in the
    $\rout$-$m_d$ plane for $m_3=1$ and $\al=1$. The contour levels
    correspond to $\log\xi = 0, -0.2, -0.4, ... , -1$ (solid lines)
    and $-1.2, ..., -1.8$ (dotted lines).  The white space on the left
    corresponds to the NDAF zone, where $\log\xi=0$.}  \label{Fig2}
    \end{center}
\end{figure}

Figures \ref{Fig1} and \ref{Fig2} show results corresponding to a
simple model in which we assume that the accretion disk is formed
instantaneously, e.g. by the disruption of a companion star.  We
specify the initial state of the disk by giving its mass $m_d$ and
initial radius $\rout$.  For such a model, we see that the region of
parameter space where NDAFs form corresponds to short accretion times
$\tacc$ of order a few tenths of a second.  This suggests that a
binary-disruption-based accretion model is capable of producing short
GRBs.  It is, however, very hard to see how such an accretion system
could produce a long burst.

One way to make a long burst from an NDAF is to decrease the value of
the viscosity parameter $\alpha$ (see eqs. 29 and 32).  Could $\alpha$
be significantly lower, say 0.01?  MHD simulations of thin accretion
disks generally give $\alpha$ in the range 0.01 upwards, and it is
widely agreed that the values obtained are lower limits since the
simulations have limited spatial resolution.  Empirical estimates of
$\alpha$ in cataclysmic variables (obtained by comparing observations
of dwarf nova outbursts with model predictions) give $\alpha\sim0.1$
in the high state (which is most relevant for our models).  There is
clear evidence that $\alpha$ is smaller, perhaps $\sim0.01$, in the
low state of CVs.  This is probably the result of the cold gas
becoming neutral and losing its coupling to the magnetic field (Gammie
\& Menou 1998), which is clearly not relevant for our ultra-hot
plasma.  We feel that $\alpha\sim0.1$ is a reasonable estimate for the
viscosity parameter.

Sakimoto \& Coroniti (1981) suggested that $\alpha$ may be lower in
radiation-dominated gases because the magnetic pressure may be in
equipartition with only the gas pressure rather than the total
pressure.  Their proposal requires that the gas and the radiation be
able to slip past each other (cf. Blaes \& Socrates 2001).  At the
extraordinarily large radiation optical depths found in our models
(both CDAFs and NDAFs), the gas and the radiation are extremely
tightly coupled.  It is therefore very unlikely that the value of
$\alpha$ would be modified.

The models we consider have large disk masses.  The accreting gas may
therefore develop gravitational instabilities (see the discussion near
the end of \S3) and lose angular momentum via gravitational waves (see
Bonnell \& Pringle 1995 and references therein).  This might lead to
an increase in the effective value of $\alpha$.  Another potential
uncertainty is that, for some choices of the parameters, electron
degeneracy becomes important.  It is not understood how the
Balbus-Hawley instability (which is thought to produce the shear
stresses behind $\alpha$) behaves in such a gas.

While the time scale of the accretion $\tacc$ depends fairly
sensitively on $\alpha$, the size of the NDAF zone is insensitive to
$\alpha$ (cf eq. 20).  Thus, Figs. 1 and 2 give fairly reliable limits
on the radius inside which the mass needs to be introduced if we wish
to have an NDAF.  We may use this information to deduce a few
interesting results on GRB models.

NS-NS and BH-NS merger models, with $(\rout, m_d)=$ (10,0.1) and
(10,0.5) (see Popham et al. 1999), are well inside the NDAF zone and,
according to our calculations, are capable of producing GRBs.
However, this is only if the black hole is small (few $M_\odot$). If
the black hole is larger than $\sim 10M_\odot$ its Schwarszchild
radius becomes too large and there is not enough ``room" for an NDAF
solution around it.  Moreover, the neutron star in this case is
swallowed whole by the BH and it is not tidally disrupted to create an
accretion disk.  On the other hand, as already noted, unless the
viscosity is much smaller than what we have assumed (which we consider
unlikely), such disks cannot produce long bursts lasting hundreds or
even tens of seconds. This suggests that NS-NS mergers and BH-NS
mergers with smallish BH masses produce the class of short duration
GRBs, but not the long duration GRBs.

Other merger models, specifically the BH-WD and the BH-He star merger
models, would appear not to be viable GRB engines.  As the secondaries
in these systems are not compact, they would form accretion flows with
large values of $\rout$.  For instance, Popham et al. (1999) estimate
$\rout\sim3000$ for a BH-WD binary and $\rout\sim5000$ for a BH-He
star binary.  At these radii, the accretion flow will be a very
exteneded CDAF and hardly any mass will be accreted.  Although the
time scales of these models are consistent with long bursts, the
extremely small value of $\macc$ suggests that these models do not
produce GRBs of any kind. It is interesting to speculate what kind of
observable events these binaries might produce (as undoubtedly they do
merge in nature).

All of the discussion so far is concerned with binary mergers, where
we have imagined that a certain fixed amount of mass is
instantaneously input into the accretion flow.  The popular collapsar
model (MacFadyen \& Woosley 1999) corresponds to a different scenario
in which mass is steadily fed over a period of time by fallback from
the supernova explosion.  MacFadyen and Woosley (1999) show that the
time scale of the GRB is set by the physics of fallback rather than by
accretion.  Further, the time scales they obtain are consistent with
observations of long GRBs.

While the time scale may be set by fallback, the efficiency of the
burst still depends on the nature of the post-fallback accretion.  Let
us ssume that fallback supplies mass at a certain injection rate
$\mdinj$ at a characteristic radius $\rout$; the latter depends on the
specific angular momentum of the material.  Figure 3 shows the
numerical results.  As expected, efficient accretion, where most of
the fallback material reaches the black hole, is possible only if
$\rout$ is small and falls within the NDAF zone.  If collapsars have a
distribution of $\rout$, then our calculations suggest that only those
systems that have $\rout\ \sles\ 100 \alpha_{-1}^{-2/7} m_3^{-1}$ will
make bursts.  Systems with larger specific angular momentum, and hence
larger $\rout$, will form CDAFs and will eject most of the mass. Such
systems may make very interesting supernova explosions, but if at all
they make GRBs the bursts are likely to be very weak.

We have assumed in this paper that the energy of a GRB is proportional
to the total mass accreted on the central compact object. For NS-NS
and NS-BH binaries, which give rise to short duration bursts, the disk
mass is expected to be nearly constant and hence the total energy
might be roughly the same for all bursts. However, for long bursts,
which we assume result from the collapse of a massive star, the energy
release depends on the angular momentum of the stellar core (which
determines $\rout$) and the mass of the stellar envelope that falls
back on the collapsed core. Observations of long duration bursts
indicate that the energy does not vary much from one burst to another
(Panaitescu and Kumar 2001, Frail et al. 2001).  This means that, for
some reason, the total accreted mass on the central
object depends only weakly on the properties of the progenitor star.

In our calculations we neglected photo-disintegration of nuclei, which
Popham et al. (1999) and MacFadyen \& Woosley (1999) show to be an
important coolant of the accreting gas.  If we include this effect,
the boundary between CDAFs and NDAFs will move to somewhat higher
values of $\rout$, perhaps by a factor of a few.  However, it will not
change our key conclusions.

We are grateful to Eliot Quataert and Andrew MacFadyen for useful
discussions, and Martin Rees for comments on the manuscript.  We thank
the organizers of the Jerusalem Winter School in Physics for
hospitality while part of this research was carried out.  This work
was supported in part by NSF grant AST 9820686 and by a US-Israel BSF
grant 98-00225.

\vfill\eject
\begin{appendix}
\section{Mass Accretion versus Ejection in CDAF Models}

We provide an alternate ``derivation'' of equation (10), and discuss
the relevant uncertainties in the result.

In a CDAF, there is a flow of energy from the inside to the outside as
a result of convection.  Let us write the convective luminosity as
$$
L_c=\epsilon_c\mdin c^2,
$$
where $\mdin$ is the mass accretion rate onto the black hole.  Since
the convective luminosity is proportional to $\alpha$ (Quataert \&
Gruzinov 2000; Narayan et al. 2000), we may write
$\epsilon_c=\alpha\eta$, with $\eta$ roughly a constant.  Ball et
al. (2001) computed a global model of a CDAF with $\alpha=0.03$, and
found that $\epsilon_c=0.0045$.  This suggests that $\eta\approx
0.15$.

In our GRB model, there is practically no radiative emission from the
gas.  Therefore, the entire convective luminosity $L_c$ must be
converted into thermal and mechanical energy of the gas on the
outside.  This energy will cause the gas at $r\sim\rout$ to move to
larger radii.

How much energy do we need to evaporate gas from $r=\rout$?  Since the
gas in a CDAF is hardly bound to the black hole --- the Bernoulli
parameter is close to zero or even positive (cf Quataert \& Gruzinov
2000; Narayan et al. 2000) --- very little energy is needed to drive a
mass outflow to infinity.  Let us write the energy carried away per
unit mass of escaping gas as $\zeta GM/\rout R_S$.  The rate at which
mass is ejected is then given by
$$
\mdout={L_c\rout R_S\over \zeta GM}={0.2\eta\al\over\zeta}\rout\mdin.
$$
Thus we have
$$
{\macc\over m_d}={\mdin\over\mdout}={5\zeta\over\eta}\al^{-1}\rout^{-1}.
$$
Equation (10) of the main paper (derived by a different approach) has
the same scaling, but with a coefficient equal to 14.

By the Bernoulli parameter argument mentioned earlier, we expect
$\zeta$ to be significantly less than unity; we would arbitrarily
guess that $\zeta\sim0.1-0.3$.  Thus we expect $5\zeta/\eta\sim10$
(compared to 14 in the main text).  However, the coefficient is
uncertain since $\zeta$ could be much smaller than our assumed value,
or conceivably much larger.  In the latter case, the convective
luminosity must be carried away by a small amount of gas that is
accelerated to a speed much greater than the escape velocity. This
possibility is not supported by the numerical simulations carried out
to date.  Our estimate of $\eta$ is also uncertain since it is based
on a single global model calculated by Ball et al. (2001).  A better
estimate could be obtained from full-scale numerical simulations.

The simulations reported by Igumenshchev \& Abramowicz (2000) may be
used to obtain a direct estimate of $\mdin/\mdout$.  For their model L
($\gamma=4/3$, $\alpha=0.03$, $\rout\sim7000$), they estimate
$\mdin/\mdout\sim0.003$, whereas equation (10) of the present paper
predicts a value of 0.007.  The two estimates agree to within a factor
of about 2, which is reassuring.  On the other hand, for $\alpha=0.1$
(Model I), they obtain $\mdin/\mdout\sim 0.02$ which is much larger
than our formula would predict.  A possible explanation is that Model
I does not strictly correspond to a CDAF.  The flow appears to be a
transition case where turbulent convection is replaced by a large
scale circulation.  In the case of MHD simulations (Stone \& Pringle
2000) the mass outflow rate is again found to be much larger then the
mass accretion rate on the central object.  Therefore, the main
results discused here should still be valid.

\end{appendix}

\begin{figure}
  \begin{center}
    \includegraphics[width=12cm]{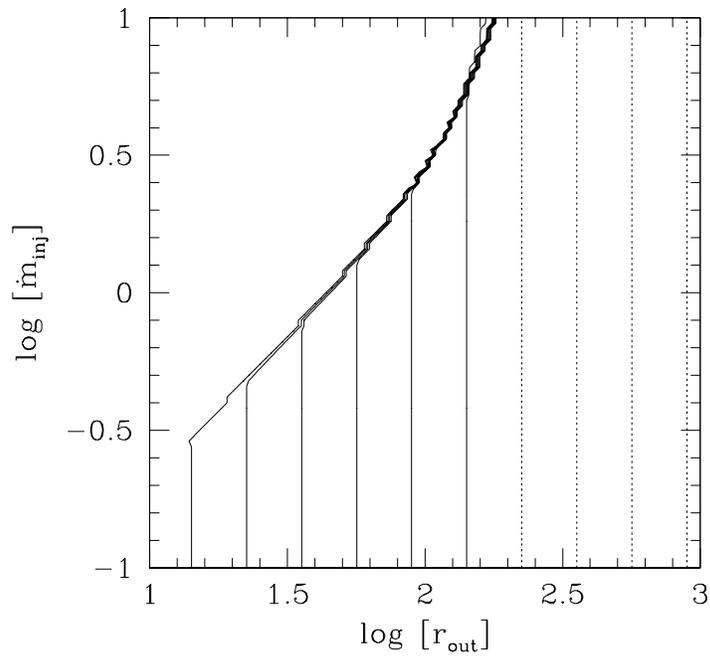}
    \caption{Contours of accretion efficiency $\xi=\md/\mdinj$ in the
   $\rout$-$\mdinj$ (in solar masses per second) plane.  The contour
   levels and spacing are the same as in Fig. 2.}  \label{Fig3}
   \end{center}
\end{figure}


\newpage

\begin{thebibliography}{99}



\bibitem[Abramowicz et al. (1995)]{A95} Abramowicz, M., Chen, X.,
Kato, S., Lasota, J.P., \& Regev, O., 1995, ApJ 438, L37

\bibitem[Abramowicz et al. (1988)]{A88}
Abramowicz, M., Czerny, B., Lasota, J.P., \& Szuszkiewicz, E., 1988, ApJ 332,
 646

\bibitem[Balbus \& Hawley, (1991)]{BH91}
Balbus, S. A. \& Hawley, J. F.
 1991, ApJ, 376, 214.

\bibitem[Ball, Narayan \& Quataert, (2001)]{BNQ01}
Ball, G., Narayan, R., and Quataert, E., 2001, to appear in ApJ,
   astro-ph/0007037

\bibitem[Begelman (1978)]{B78}
Begelman, M., C., 1978, MNRAS 243, 610

\bibitem[Blaes \& Socrates, (2001)]{BS01}
Blaes, O., \& Socrates, A., 2001, ApJ, submitted (astro-ph/0011097)

\bibitem[Bloom et al. (2000)]{Bloom00}
Bloom, J.S., Kulkarni, S.R., and Djorgovski, S.G., 2000, AJ,
submitted (astro-ph/0010176)

\bibitem[Bonnell \& Pringle, (1995)]{BP95}
Bonnell, I. A., \& Pringle, J. E., (1995), MNRAS, 273, L12

\bibitem[Chen et al. (1995)]{Chen95}
Chen, X., Abramowicz, M., Lasota, J.P., Narayan, R., and Yi, I., 1995, ApJ
  443, 61

\bibitem[Eichler et al., (1989)]{Eichler89}
Eichler, D. Livio, M.; Piran, T. \& Schramm, D. N., 1989, Nature, 340, 126.

\bibitem[Frail et al. (1997)]{Frail97}
Frail, D. A., Kulkarni, S. R., Nicastro, L., Feroci, M., \& Taylor, G. B. 1997,
Nature, 389, 261

\bibitem[Frail et al. (2001)]{Frail01}
Frail, D.A. et al., 2001, Nature, submitted (astro-ph/0102282)

\bibitem[Frank, King \& Raine (1992)]{FKR92}
Frank, J.; King, A.; Raine, D., 1992,  Accretion Power in Astrophysics,
Cambridge Univ. Press

\bibitem[Fryer \& Woosley, (1998)]{FS98}
Fryer, C. L.; Woosley, S. E. 1998, ApJ, 502L.

\bibitem[Fryer et al., (1999)]{FWHD99}
Fryer, C.L.; Woosley, S. E.; Herant, M.; Davies, M.B. 1999, ApJ,520,650.

\bibitem[Gammie \& Menou, (1998)]{GM98}
Gammie, C. F., \& Menou, K., 1998, ApJL, 492, L75

\bibitem[Igumenshchev et al. (1996)]{I96}
Igumenshchev, I.V., Chen, X., and Abramowicz, M.A., 1996, MNRAS 278, 236

\bibitem[Igumenshchev \& Abramowicz (1999)]{I99}
Igumenshchev, I.V., and Abramowicz, M.A., 1999, MNRAS 303, 309

\bibitem[Igumenshchev \& Abramowicz (2000)]{I00}
Igumenshchev, I.V., and Abramowicz, M.A., 2000, ApJS 130, 463

\bibitem[Igumenshchev et al. (2000)]{I00a}
Igumenshchev, I.V., Abramowicz, M.A., and Narayan, R., 2000, ApJ 537, L271

\bibitem[Kato et al. (1998)]{K98}
Kato, S., Fukue, J., \& Mineshige, S., 1998, in "Black-hole accretion disks",
Edited by Shoji Kato, Jun Fukue, and Sin Mineshige;
                   Publisher: Kyoto University Press


\bibitem[Katz (1977)]{K97}
Katz, J., I., 1977, ApJ 215, 265

\bibitem[Katz \& Piran (1997)]{KP97} Katz, J., I., \& Piran, T.,
1997, ApJ 490, 772

\bibitem[MacFadyen \& Woosley (1999)]{MCW99}
MacFadyen, A. I. \& Woosley, S. E., 1999, ApJ, 524,262

\bibitem[Nakar \& Piran (2001a)]{NP01a}
Nakar, E., \& Piran, T., 2001a, MNRAS, submitted (astro-ph/0103192)


\bibitem[Igumenshchev et al. (2000)]{NIA00}
Narayan, R., Igumenshchev, I.V., and Abramowicz, M.A., 2000, ApJ 539, 798

\bibitem[Narayan et al. (1998)]{NMQ98}
Narayan, R., Mahadevan, R., and Quataert, E., 1998, "The Theory of Black Hole
      Accretion Discs", eds. M. A. Abramowicz, G. Bjornsson, and J. E. Pringle

\bibitem[Narayan, Paczynski \& Piran (1992)]{NPP92}
Narayan, R., Paczynski, B., \& Piran, T., 1992, ApJ, 395L, 83

\bibitem[Narayan, \& Yi (1994)]{NY94}
Narayan, R., and Yi, I, 1994, ApJ 428, L13

\bibitem[Narayan, \& Yi (1995)]{NY95}
Narayan, R., and Yi, I, 1995, ApJ 444, 231

\bibitem[Paczynski, (1991)]{Pac91}
Paczynski, B. 1991, Acta Ast, 41, 257

\bibitem[Paczynski, (1998)]{Pac98}
Paczynski, B., 1998, ApJ, 494L, 45

\bibitem[Panaitescu \& Kumar, (2001)]{PK01}
Panaitescu, A., and Kumar, P., 2001, ApJ, in press (astro-ph/0010257)

\bibitem[Piran (1999)]{P78} Piran, T., 1978, ApJ 221, 652

\bibitem[Piran (1999)]{P99} Piran, T., 1999, Phys. Reports, 314, 575

\bibitem[Piran (2000)]{P00} Piran, T., 2000, Phys. Reports, 333, 529

\bibitem[Popham, Woosley and Fryer, (1999)]{PWF99}
Popham, R.; Woosley, S. E.; Fryer, C., 1999, ApJ, 518, 356

\bibitem[Quataert \& Gruzinov (2000)]{QZ00}
Quataert, E., and Gruzinov, A., 2000, ApJ 539, 809

\bibitem[Ruffert \& Janka (1999)]{RuffertJ99}
Ruffert M., \& Janka H.-Th., 1999, A \& A, 344, 573

\bibitem[Sakimoto \& Coroniti, (1981)]{SC81}
Sakimoto, P. J., \& Coroniti, F. V., 1981, ApJ, 247, 19

\bibitem[Sari \& Piran, (1997)]{SP97}
Sari, R.,  Piran, T., 1997, ApJ, 485, 270.

\bibitem[Shakura \& Sunyaev, (1973)]{SS73}
Shakura, N. I.; Sunyaev, R. A. 1973, A\&A, 24, 337.

\bibitem[Stone et al. (2000)]{Stone00}
Stone, J.M., and Pringle, J.E., 2000, MNRAS 322, 461

\bibitem[Stone et al. (1999)]{Stone99}
Stone, J.M., Pringle, J.E., and Begelman, M.C., 1999, MNRAS 310, 100

\bibitem[Usov, (1992)]{Usov92}
Usov, V. V. 1992, Nature, 357, 472.

\bibitem[Woosley, (1993)]{Woosley93}
Woosley, S. E. 1993, ApJ, 405, 273.


\end{thebibliography}
\end{document}